# Deaf in AI: AI language technologies and the erosion of linguistic rights

**Maartje De Meulder[1]**
**HU University of Applied Sciences, Utrecht**

Abstract

This paper explores the interplay of AI language technologies, sign language interpreting, and linguistic access, highlighting the complex interdependencies shaping access frameworks and the trade-offs these technologies bring. While AI tools promise innovation, they also perpetuate biases, reinforce technoableism, and deepen inequalities through systemic and design flaws. The historical and contemporary privileging of sign language interpreting as the dominant access model – and the broader inclusion ideologies it reflects – shape AI's development and deployment, often sidelining deaf languaging practices and introducing new forms of linguistic subordination to technology. Drawing on Deaf Studies, Sign Language Interpreting Studies, and crip technoscience, this paper critiques the framing of AI as a substitute for interpreters and examines its implications for access hierarchies. It calls for deaf-led approaches to foster AI systems that remain equitable, inclusive, and trustworthy, supporting rather than undermining linguistic autonomy and contributing to deaf-aligned futures.

## Introduction: AI, access, and the erosion of linguistic rights

After a recent presentation I gave on ethical considerations in sign language AI, a European deaf attendee approached me. Like me, they had an advanced degree, high socioeconomic status, fluency in several signed and written languages, and digital and AI literacy, along with reliable access to human sign language interpreting services. They were also a public figure engaged in interpreting and linguistics research. With visible frustration, they confided that they were fed up with the current state of human sign language interpreting services in their country and beyond – the logistical hurdles, the pervasive audism, the loss of privacy, and the economics of interpreting services. "AI", they said, "might finally free us from this system. Maybe in the not-so-distant future, we won't need human interpreters anymore." They paused, then added with a wry smile, "Of course, saying that out loud would cause a storm."

[1] HU University of Applied Sciences Utrecht, The Netherlands; Faculty of Healthy and Sustainable Living, Research Group Speech and Language Therapy: Participation through Communication, Heidelberglaan 7, 3584 CS Utrecht. ORCID ID: 0000-0001-7607-5314

Yet, they were not the first person (of a comparable position) telling me this, and I assume they will not be the last. In a recent paper U.S. deaf authors Hall, Brick, and Millios (2024:10) state: "It must be acknowledged that if Deaf people could go without interpreters, they would." Those deaf people involved in research on interpreting, working as interpreters, and/or being on the receiving end of interpreting services are aware of the limitations of human interpreting in terms of providing equitable "access" (Haualand, De Meulder, & Napier, 2023). For some, AI tools such as speech-to-text applications may offer a (partial) relief from the access labour associated with working with human sign language interpreters. At the same time, deaf communities are generally acutely aware of how access is collective and interconnected, recognising that access is not just about personal choices but about shared opportunities and systemic structures.

There is also an implicit deaf understanding of access rights as, what I call, "slippery slope rights", referring to measures initially rejected by deaf communities due to their limitations, only to be normalised over time as acceptable solutions. For instance, while deaf people have objected against the use of Video Remote Interpreting (VRI) – a service that connects users to remote interpreters through video technology – in hospitals because of unreliable technology, issues with establishing trust, and privacy, the use of VRI in some medical situations is now considered acceptable by health care providers (James et al., 2022; Kushalnagar et al., 2019).

The attendee who approached me knew there is a difference between saying (like Hall, Brick, & Millios, 2024) "if we could do without interpreters, we would" or saying out loud that AI could be a replacement for or even serve as a tentatively acceptable addition to human interpreters. They knew very well this could trigger governments and institutions to abandon interpreter services altogether, in favour of "cheaper" AI-driven solutions. For many deaf people without similar privileges – those with lower AI literacy, less access to reliable AI tools, lower socioeconomic status, and/or restricted access to human sign language interpreting services (the reality for most deaf people around the world) – this shift could have devastating consequences, potentially eroding hard-won linguistic and accessibility rights. The implementation of AI technologies risks exacerbating access hierarchies and shifting labour onto individuals without adequate resources or support. This raises critical questions about how these tools redistribute access and reshape social hierarchies. Deaf-aligned futures depend on ensuring deaf people retain control over their access choices, resisting the imposition of technologies that may fail to align with their preferences. While some deaf users may embrace AI tools as empowering, others who prefer working with human interpreters might face a narrowing of choices.

This paper examines this push and pull of AI tools in the context of deaf people, sign language interpreting, and linguistic access, highlighting the complex deaf interdependencies that underpin access frameworks, the trade-offs inherent in these technologies, and the critical need to navigate these shifts carefully. Because in practice, the erosion of linguistic rights may already be underway. Like many technological shifts, this does not happen overnight but gradually, through subtle shifts in policy, funding priorities, and service provision. These incremental changes often remain unnoticed until they become deeply embedded in systems, leaving those most reliant on access services with fewer options and greater disparities. I explored some initial thoughts on this in an earlier article (De Meulder, 2021). This was just before the rapid rise of generative AI and its uptake by everyday users, which has significantly shifted the landscape. Today, I am writing in a constantly and quickly evolving environment, as AI capabilities advance, and the role of sign language interpreters evolves.

These rapid shifts are reshaping the very fields they affect, challenging established assumptions, theories, and practices. Firstly, AI disrupts long-standing models in Sign Language Interpreting Studies and Deaf Studies that primarily centre human interpreters and human-to-human interaction as the dominant means of access and mediation (Napier, McKee, & Goswell, 2010; Stone et al., 2022; Young, Napier, & Oram, 2019). With some exceptions (e.g. Clark, 2021), these theories have yet to fully address the implications of these developments. Secondly, AI has significant implications for the theoretical and methodological framework of sign language (socio)linguistics and has the potential to disrupt deeply held attitudes and ideologies about (signed) languages (Backus et al., 2023; Kelly-Holmes, 2024). With a few exceptions (e.g. Lee, Hill, & Smith, 2021), research on sign language ideologies and sign language sociolinguistics so far has focused exclusively on human-to-human interactions, relying on established theoretical and methodological frameworks (Kusters et al., 2020). These fields must now grapple with the implications of human-AI interactions, including how these influence perceptions of access, mediation, interpreting, authority, correctness, and acceptability in language use.

## Setting the stage

The widespread adoption of AI language technologies by everyday users is reshaping how we communicate, work, and engage with the world. Generative language AIs, powered by Large Language Models (LLMs), understand, process and (co-)produce human-like language across various contexts. Current examples of these technologies include tools like Google Translate and automated captions on platforms such as Zoom, intelligent personal assistants like Siri and Alexa, chatbots, and speech-to-text applications such as Otter.ai. More advanced conversational AIs, such as ChatGPT and Gemini, further demonstrate the growing sophistication of LLMs. Beyond text, AI's generative capabilities extend to the visual domain with tools like DALL-E and Midjourney, producing images outputs from textual prompts. At the same time, AI systems can interpret visual input in real-time, describing objects, environments, or scenes through spoken or written descriptions.

In the not-so-distant future, these tools are expected to include specialised sign language technologies – those designed for sign language recognition, generation and translation. Examples of these technologies are virtual sign language avatars and automated translation interfaces between signed and spoken languages. The barriers to developing sign language technologies (primarily the volume and quality of data) have been well-documented (De Coster et al., 2023; Desai et al., 2024; Shterionov, Leeson, & Way, 2024; Vandeghinste et al., 2024; Yin et al., 2024). While these challenges mean that these technologies currently remain underdeveloped for broad use, the state-of-the art is advancing rapidly. As I write this in January 2025, another new project "SignGPT"[2] (£8,45m) has just been announced, aiming to build a British Sign Language LLM capable of end-to-end translation English/BSL and build "the largest sign language dataset in the world". Reflecting broader trends in the field of sign language AI (Desai et al., 2024), all the Principal Investigators are hearing academics, and the data (also) include interpreted datasets English/BSL (Albanie et al., 2021). I will come back to these trends and biases later in the article.

These AI language technologies are evolving into semi-autonomous communicative agents that mimic 'human' forms of interaction. Important in the context of sign languages and deaf people is that these tools operate across various modalities (text, speech, gestures, movement, sign) and take on different degrees of embodiment, from text-based interfaces to human-like robots and virtual avatars. Yet, the implications of these AI technologies for deaf people (who use sign languages and navigate diverse communication practices) remain largely unexplored or primarily examined from hearing, non-disabled perspectives, although there are emerging exceptions (e.g. Angelini et al. in preparation; Glazko et al., 2023; Huffman et al., 2024).

[2] https://www.surrey.ac.uk/news/signgpt-project-awarded-ps845m-build-sign-language-ai-model-deaf-community

In language policy and legislation, deaf people are often positioned as both linguistic minorities and individuals with disabilities (De Meulder & Murray, 2017). This dual status grants deaf people (primarily in Europe, Australia, New Zealand, Canada and the United States) access to sign language interpreting services. While there are parallels between public service interpreting for deaf people and other groups who use interpreters for accessing public services such as immigrants, refugees, and other language groups (Stone, 2010), there are also differences. For immigrants and refugees, the demand for interpreting services often fluctuates with migration patterns, which means the request for interpreting in certain languages rises and falls over time (Giambruno, 2014; Piller, 2017; Skaaden & Wadensjö, 2014). For immigrants, interpreting services are frequently framed as a temporary measure, intended to bridge the gap until they acquire proficiency in the majority language. This creates an intersecting dynamic for deaf migrants and refugees (Duggan & Holmström, 2024; Sivunen, 2019). The perceived temporality of spoken language interpreting services is shaped by discourses of who is a deemed a 'worthy' recipient of interpreting, particularly within assimilationist institutional practices that enforce the learning of the majority language (Piller, 2017). By contrast, sign language interpreting services are generally provided over deaf people's lifetime and are not limited to the public sector, as if often the case for spoken language interpreting, but provided across a wide range of domains. For example, in several Western countries, a significant portion of sign language interpreters work in educational settings, including early childhood and primary education, domains where spoken language interpreters tend not to work. This divergence reflects deaf people's categorisation as disabled and the associated ideologies of inclusion, which justify placing young deaf children in regular education settings with interpreters. This practice, and the larger practice of sign language interpreting services to fulfil ideologies of inclusion has been criticised as an "illusion of inclusion", masking deeper systemic inequities (Caselli, Hall, & Henner, 2020; De Meulder & Haualand, 2021; De Meulder & Murray, 2021; Russell, 2021).

This intersection of technological progress and systemic challenges creates a critical moment of both ruptures and opportunities for resilience, with profound implications for access and communication practices, as well as for theory-building in the field. How do these dynamics unfold in practice? To answer this question, I will turn to how "access" is conceptualised in Deaf Studies, Sign Language Interpreting Studies, and crip technoscience, providing a framework to critically evaluate discourses surrounding the development and use of AI (sign) language technologies. Indeed, the issue of 'access' (to information, education, political participation, employment, health care, the justice system, etc.) has been paramount on international deaf agendas, having been the central advocacy issue for national, regional and international deaf NGOs for many years, and takes a central place in discourses around the development and use of AI language technologies.

Before proceeding, let us first have a look at the complex interplay between AI-driven language technologies and human sign language interpreting services in deaf lives, and how these systems interact in everyday practice. Consider a typical day of a literate deaf person with a high socioeconomic status (like me) in 2025: a vibrating smartwatch wakes them, real-time captions generated by automatic speech recognition relay the news. Their hearing family members ask Alexa for weather updates through voice commands – it does not support sign language yet. At work, Zoom meetings offer live captions alongside a remote human sign language interpreter, but the captions often misrepresent jargon and the interpreter, unfamiliar with the topic, struggles to keep up. During lunch, a call to the doctor involves a remote human interpreter via a Video Relay Service. While effective in mediating the conversation, the reliance on a third party to access private health information feels intrusive. Later, a visit to the doctor's office prompts a different choice: rather than using an interpreter, they opt for a speech-to-text app, which feels more discreet. Yet, the app's accuracy is inconsistent, creating moments of uncertainty, misunderstanding or just plain awkwardness. At a local shop, they use a speech-to-text app to communicate with the shop assistant. The app struggles with environmental noise and regional accents, turning a simple low-stake conversation into a prolonged source of frustration. For a parent-teacher meeting they request an interpreter since the teacher is not comfortable speaking into a phone and the WIFI at the school is spotty. However, the school hires an unfamiliar interpreter, leading to a strained conversation. Later in the day they attend a court hearing as part of a dispute over workplace discrimination. Although they requested a qualified interpreter, the court arranges for an AI-powered translation instead, citing efficiency and cost-saving measures.

This vignette illustrates how AI-driven language technologies and human sign language interpreting services interact in shaping deaf communication and access for deaf people with high literacy and socioeconomic status privileges. The outcomes would differ significantly for deaf people without those privileges, i.e. without the same technology, legislation, internet access, etc. Moreover, as AI capabilities advance and the role of interpreters evolves, a vignette written in a few years' time could sketch an entirely different landscape.

**The intersection of AI sign language technology development and ideologies centred on interpreting**

The historical and contemporary situation of sign language interpreting services and the associated discourses around 'access' significantly influence how AI language technologies are being developed and (will be) deployed in this domain. In Europe, Australia, New Zealand, Canada and the United States, sign language interpreting services have now existed for more than four decades and have evolved from volunteer charity work to well-established social institutions and professional services. However, the availability of sign language interpreters is often equated with access, with increasing the number of interpreters posited as the primary solution to access issues. This perspective is so entrenched that interpreter services are now a prerequisite for public service provision (De Meulder, Murray, & McKee, 2019; De Meulder & Haualand, 2021). Yet training sign language interpreters, who predominantly represent white, female, hearing, abled, new signer demographics – and thus become a mostly white, female, hearing, abled profession (Hill, Isakson, & Nakahara, 2022; Napier et al., 2022) requires substantial time and resources. Virtually all sign language interpreter training programs struggle with teaching sign language acquisition *and* the acquisition of subject-specific knowledge within a three or four-year (BA or MA) program (Webb, Napier, & Adam, 2025). As a result, interpreter education and provision represent a limited human capacity solution.

Simultaneously, in most of these countries, institutionalised sign language interpreting services are undergoing significant shifts driven by multiple converging forces. They face logistical and capacity constraints, challenging their sustainability as a means of providing access. These challenges manifest differently across contexts: for example, in Sweden, a retirement wave is creating shortages (DIK, 2024), while Finland's saturated labour market raises concerns over interpreters' job security (Rainò, 2022). In Germany, the shortage of sign language interpreters creates tensions between (hearing) sign language interpreters and deaf consumers (Maaß, 2024). In more general terms, the pressures stem from systemic supply-demand imbalances, ideologies of inclusion that over-rely on human interpreters, and an aging workforce outpacing new entrants. Simultaneously, in many of the same countries that experience these pressures, some deaf people are achieving higher levels of education and literacy in written languages, especially in professional and academic sectors. This, and the rapid uptake of generative AI, also by deaf users, may be changing reliance on human interpreters to meet access needs, prompting shifts in how and when these services are sought.

These dynamics are compounded by perceptions of governments and other policy actors as interpreting services as logistically complex and costly, contributing to a potential growing push towards AI-driven solutions as a more 'cost-effective' alternative. My own work (De Meulder & Haualand, 2021), highlighting the limits of sign language interpreting as a quick fix for inclusion, is regularly cited to motivate the development of AI sign language tools (e.g. Coy et al., 2024; Gunarhadi et al., 2024; Nasution & Batubara, 2024). Anecdotal evidence from some European countries indicates that deaf (young) adult consumers of interpreting services are increasingly directed to rely on speech-to-text AI (also in education settings) when their allotted interpreting hours are exhausted or when human interpreters are unavailable. This shift highlights the slippery slope of access rights and raises concerns about choice and agency.

A related issue I preliminary discussed in earlier work (De Meulder, 2021) is that the primary bottleneck for developing AI sign language technologies is and remains the scarcity and quality of available data. Here again, sign language interpreting plays a significant role. AI systems are inherently data-hungry and due to the lack of data "in the wild", developers often turn to readily available internet data, including interpreted data sets from, among other things, public broadcasting. Most language AI uses language pairs to learn, so in theory this is not uncommon. What is uncommon is that the datasets are not just language pairs, but *simultaneously interpreted* language pairs. These datasets include both the signed input from (mostly hearing, mostly new signer) interpreters and the spoken source languages, or vice versa, with the signed source language from for example deaf presenters or interpreters paired with spoken output. The use of interpreted datasets as source material to develop sign language AI has been acknowledged as problematic (e.g. by Desai et al., 2024; Fox, Woll, & Cormier, 2023; Vandeghinste et al., 2024). Indeed, interpreted language use differs significantly from usage patterns in deaf communities and the nature of scripted and interpreted language use may result in a distorted representation of sign languages in AI systems. However, this reliance on interpreted datasets not only skews representations of sign languages in AI systems but also reflects and reinforces broader discourses that prioritize interpreting as the dominant framework for accessibility in both policy and practice.

National and international deaf and sign language interpreter associations still largely view sign language interpreting as the primary or preferred method of access, particularly in what they call "high-stakes contexts" such as medical settings, educational settings, live broadcasting, and legal proceedings (WFD & WASLI, 2018). This is confirmed by recent data from interviews with deaf users in Europe, who list the same settings as high-stakes. Although, they further fine-tune them depending on whether the information flow is one-way (AI) or two-way (interpreter), short (AI) or long (interpreter), with high (interpreter) or low (AI) impact, planned in advance (interpreter) or of a fleeting, spontaneous nature with low-impact (AI) (Picron, Van

Landuyt, & Omardeen, 2024a). Current scholarly and practice discourses (e.g. Picron et al., 2024b) frequently compare AI to human interpreters, emphasising that the quality of human interpreters is, and will likely remain, superior to AI-driven alternatives. Research often frames AI and human interpreters within a hierarchy, with human interpreters consistently positioned as superior (SignOn Consortium, 2023). This framing, taken together with the use of interpreted datasets, perpetuates a problematic benchmark for AI sign language technologies.

Indeed, these technologies are not meant to substitute or compete with interpreters; they are meant to support and reflect deaf languaging practices, i.e. the ways deaf people themselves use and navigate signed and written languages in different contexts. Furthermore, bias risks are amplified when AI systems are trained on an amalgams of data sets containing interpreted input. This concern is heightened by the already significant influence of interpreters on deaf people's signing. For example, sign language interpreters often serve as de facto language models for deaf learners in regular educational settings (Caselli, Hall, & Henner, 2020), and deaf people frequently adjust their signing to facilitate comprehension by interpreters. This is a form of linguistic subordination to interpreters that now risks being compounded by linguistic subordination to technologies. In my previous work (De Meulder, 2021), I cautioned against the potential use of AI sign language technologies, especially those trained on interpreted datasets, in sign language interpreter training programs. This concern remains pressing, particularly in a context of ongoing budget cuts to higher education programs and a neoliberal climate that prioritises quantity over quality, with limited resources for staff and students alike (Webb, Napier, & Adam, 2025). This would exacerbate existing problems in sign language interpreter training programs, which often already fail to reflect the racial, multicultural and multilingual diversity of deaf communities (Robinson, Sheneman, & Henner, 2020).

However, in the current climate marked by an emerging co-existence of AI language technologies with human interpreting, deaf NGOs may find themselves in a double bind. Their position is shaped not only by their commitment to ensuring reliable and nuanced access, but also by an awareness that access is inherently collective, extending beyond individual accommodations to encompass the diverse needs, preferences, and circumstances of different deaf communities. This means balancing different priorities, levels of technological accessibility, and situational constraints, while advocating for solutions that are both scalable and adaptable across various contexts and individuals. Indeed, access through human interpreters already generates hierarchies among deaf people (De Meulder & Haualand, 2021). Deaf people who can deploy multilingual and multimodal resources, often work with interpreters and have access to the "best" interpreters (because of professional experience, cooperation with interpreters, and network) seem to get more out of interpreter-mediated interactions than deaf people who do not have these privileges or resources. AI introduces additional (although possibly different) layers of inequality, arising from differential access to different AIs and varying levels of digital and linguistic literacy among deaf users. These dynamics illustrate the intersection of sign language technology development and inclusion ideologies centred on interpreting. However, there are even additional layers of complexity.

*Ambivalence and resilience in interpreting relationships*

Building on the dynamics described above, it is crucial to situate them within the broader context of issues that AI sign language technology developers (who in most cases have no lived experience of being deaf) often overlook. Deaf people have a historical and present complex and ambivalent relationship with sign language interpreters – marked by both individual critiques on interpreters and the broader interpreting system, while at the same time a reliance on this system for access. Hall, Brick and Millios (2024), whom I referred to in the introduction, state that "at a deeply fundamental level, Deaf people do not want interpreters in their lives; they would prefer to navigate the world on their own" (p. 11). This is an issue that is often not spoken about and not acknowledged by most hearing people, and many interpreters themselves. Interpreters are frequently perceived by deaf people as (un)invited guests in their personal lives (Blankmeyer Burke, 2017; Hall, Brick, & Millios, 2024) and as oppressive actors and gatekeepers unaware of their own power and privilege (Robinson, Sheneman, & Henner, 2020). For a large part, this also has to do with how they are being trained (Sheneman & Robinson, 2021; De Meulder & Stone, 2024). Yet, in Deaf Studies and Sign Language Interpreting Studies, the relationship between deaf people and sign language interpreters is often still described as being "bound together in mutual systems of precarious interdependence" (Marie, 2019 in Marie & Friedner, 2021:5), with deaf lives seen as "predicated on interpreters" and "deaf selves" largely existing in translation (Young, Oram, & Napier, 2019; Young, Napier, & Oram, 2020). Sign Language Interpreting Studies has also examined the issue of 'trust' as a crucial element in the working relationships between sign language interpreters and deaf people, with interpreters and deaf consumers often claiming that trust is a prerequisite to a successful interaction (De Meulder et al., 2018; Napier, 2011). Although, this concept of trust in interpreted interactions has also been critically examined by deaf scholars (O'Brien et al., 2023). Emerging research, again not coincidentally, mostly done by deaf scholars (Haualand, De Meulder, & Napier, 2023), has exposed the "access labour" for deaf people working with human sign language interpreters. This refers to the (often invisible) additional emotional, cognitive and logistical efforts deaf consumers are dealing with in these situations. This includes "gaze work" (De Meulder & Stone, 2024), preparing and "grooming" interpreters (De Meulder, Napier, & Stone, 2018), dealing with interpreters' insecurities, and negotiate relationships with them (Chua et al., 2022; Crawley & O'Brien, 2020). Chua et al. (2022) found that working with sign language interpreters is one of the main reasons for impostor syndrome in deaf academics.

It is therefore not surprising that some deaf people, particularly those with (high) literacy privileges, start to see AI as offering resilience: AI tools reduce the need for emotional labour, have no personal attitudes (but, biases nonetheless), reduce the need for gaze work, and are generally free of charge and accessible around the clock. Another notable form of resilience is the mainstream adoption of these technologies: hearing people are also integrating them into their practices. Many hearing people are used to speaking into a speech recognition app on a phone and a range of technologies now incorporate mainstream accessibility features such as live captions, vibrational alerts, and visual notifications such as flash alerts. Wearable technologies, such as smartwatches and emerging devices like smart glasses are gaining broader social acceptance. These devices blend function with fashion in a way that makes them "cool" (see also Profita et al., 2016). In contrast, the presence of a human interpreter is a very visible form of mediation and often carries a social stigma for deaf people, with interpreters being perceived as family members or caregivers, or their performance leading to negative assumptions about deaf peoples' capabilities (Feyne, 2018; Heldens & Van Gent, 2020). Yet, these perceived benefits must be critically examined against their trade-offs. Crip Technoscience offers a lens to do this, interrogating not just how these tools are being used but also how they redistribute access and reshape social hierarchies.

**Access and crip technoscience**

Crip Technoscience (Hamraie & Fritsch, 2019) offers a critical lens to understand the complex, often ambiguous relationship between disabled people and technology. This may mean disabled people adopt imperfect, uncomfortable, or less functional tools, while simultaneously critiquing their limitations (notice the parallels here with deaf people's ambiguous relationships with sign language interpreters). They position themselves as both users and design experts, hacking, adapting, and reimagining existing technologies to fit their desires and needs (Hamraie, 2023). Using a similar lens, recent interdisciplinary deaf-led work in Human-Computer Interaction (HCI) is exploring which technologies – and what futures – are truly *worth wanting* for deaf communities (Angelini 2024; Angelini et al., 2024a, b).

Crip Technoscience provides a critical lens for understanding deaf experiences with AI, particularly in the complex and often tense relationship between disability and technology. For example, automated captions may benefit some deaf people with strong written language skills (in dominant languages), but disadvantage others without such skills. Technologies may not work effectively in certain contexts, such as remote locations. They can lead to social inequalities between deaf people, because not all can pay for an (often more performant) AI model. Remote access can benefit some, while at the same time 'distantism' (the societal bias that prioritizes the distance senses of sight and hearing over touch) creates accessibility barriers for deafblind people (Clark, 2021). In a future projection users of larger, better institutionalized sign languages like American Sign Language (ASL) or Libras (the majority sign language in Brazil) might benefit more from sign language technologies than those sign languages with much smaller datasets available. They might also privilege some varieties of these sign languages over others, and disadvantage those deaf users with idiosyncratic signing that might influence their interaction with sign language AI.

This, and especially the combination with access (or lack of access) to human sign language interpreters, creates new hierarchies of accessibility, with "the same conditions enabling access for some can create hierarchies of belonging for others" (Hamraie, 2023, p. 310). This positions communication access as a collective entity grounded in interdependency (Bennett et al., 2018; McDonnell & Findlater, 2024). Access is co-constructed: changes in access practices, such as new technologies, often redistribute access in ways that improve opportunities for some, while potentially creating barriers for others. AI language technologies highlight these dynamics. While they may reduce stigma and access labour associated with working with human interpreters, this resilience also must be seen against the unavoidable trade-offs that accompany their adoption. These trade-offs include the fact that, like human interpreters, AI systems are prone to bias, including ableism, racism, and gender bias, which are deeply embedded in the data and algorithms that underpin them (Börstell, 2023; Broussard, 2023; Glazko et al., 2023). AI tools are also prone to hallucinating – Hicks, Humphries and Slater (2024) argue that "bullshitting" is a more fitting description – generating misleading outputs that may compromise trust. Additionally, while some of these tools are freely available, they often come with considerable ethical concerns. These include granting large corporations access to sensitive personal data and uncertainty about how this data is stored, processed and used to train models. This can lead to a lack of transparency and accountability. Furthermore, the ecological footprint and the "ghost work" involved in AI development and deployment remains a growing concern (de Vries, 2023; Gray & Suri, 2019).

Additionally, the line between resilience and breakdown is razor-thin: when the limits of technology are reached, these tools risk becoming marginalizing. For example, speech recognition apps may fail to recognise 'deaf accents', referring to the distinctive speech patterns of some deaf people, or regional dialects spoken by hearing people into a phone. In a future projection, sign language AI may encounter similar challenges in understanding regional sign languages, sign language dialects (e.g. Black ASL, which is significantly underrepresented in datasets), sociolinguistic variation (e.g. queer signing, influence of International Sign) or idiosyncratic signing styles, such as those resulting from additional disabilities, or just lack of access to signing spaces (Haualand et al., 2024). Those who can calibrate their signing to be recognised by AI sign language technologies often mirror the same group accustomed to adapting to human interpreters. This dual adaptability privileges already privileged "normative" users of well-resourced sign languages, potentially widening existing inequalities.

**Deaf leadership in sign language AI research**

These dynamics – access labour, hierarchies of accessibility, and the collective nature of access – highlight the need to study the socio-political contexts in which both these tools, and human sign language interpreting services, operate (Erdocia, Migge, & Schneider, 2024; Haualand, De Meulder, & Napier, 2023; Sayers, Sousa-Silva, & Höhn, 2021; Sergeant, 2023). Currently, the field of AI sign language is dominated by non-disabled, often non-signing, actors. While some of their work reflects systemic biases such as technochauvinism (the belief that technology is the best solution for social problems), technoableism (the drive to "fix" disabled bodies), and modality chauvinism (the privileging of speech overs sign) (Broussard, 2023; Henner & Robinson, 2023; Shew, 2023), other challenges stem from plain ignorance and working in disciplinary silos. Many researchers and developers lack meaningful engagement with deaf communities. These embedded biases shape designs like smart gloves that translate signs to text or speech (Lu, 2016). This foregrounds the belief that deaf people's ideas make more sense if converted to spoken form (Hill, 2013), or that deaf people are not worth talking to since there is no assumption of bilateral engagement, as the gloves do not translate the hearing person's speech back to sign. Similarly, AI tools are often designed with normative bodies and minds in mind: voice assistants like Alexa or Siri often fail to account for deaf accents – (Devries et al., 2024; Tran et al., 2024).

A systematic review conducted by Desai et al. (2024) based on 101 papers in AI sign language research, found that while calls for the ethical development of sign language AI (e.g. Bragg et al., 2021; De Meulder, 2021; Fox, Woll, & Cormier, 2023) have helped some individual researchers improve their practices, the field continues to face systemic biases. These include an overfocus on addressing perceived communication barriers, a lack of use of representative datasets, reliance on annotations lacking linguistic foundations, and the development of methods built on flawed models. To echo Broussard (2023), these are 'more than just glitches': these biases are now baked-in as standards, and risk causing harm by marginalising the very target users of sign language AI.

For a large part, these biases are introduced, magnified, and sustained by a lack of deaf leadership in the field of sign language AI (a problem extending to nearly every single field and discipline dealing with deaf people and sign languages). Desai et al. (2024) found that the field is mainly driven by what hearing, abled, researchers perceive as the most urgent problems and by decisions that they perceive as the most convenient. This also significantly impacts on how co-creation and co-design are filled in. While these concepts have become central in technology development and HCI, both in research and in policy (cf. the EU 2024 AI Act and 2019 Ethics Guidelines), they often appear to function more as strategies to secure funding than as genuinely well-integrated principles in research design and development. "Co-creation" often serves as user studies or data collection methods, without fundamentally challenging the status quo for disabled people. In some cases, they verge on exploitation (De Meulder, Van Landuyt, & Omardeen, 2024; Raman & French, 2022) or are perceived as "participation washing" (Sloane et al., 2020), referring to the superficial or performative involvement of participants in these processes. In large research consortia focused on AI and people with disabilities, I often find myself to be the only disabled researcher in the room. There is a growing acknowledgment of this under-representation of deaf researchers in the field. Shterionov, Leeson & Way (2024:18) acknowledge the "collective responsibility to ensure that we create pathways to the field that welcomes, supports, and encourages deaf scholars. Indeed, ideally, deaf researchers would be leading work on SLMT". However, this underrepresentation is often acknowledged in a detached manner, with researchers presenting it as an unfortunate reality rather than a system issues that requires their active intervention. Many projects, including those emphasising co-creation, continue to be funded without deaf Principal Investigators, while the predominant response remains one of lamenting the lack of deaf researchers rather than addressing the structural barriers that contribute to this lack. Moreover, the growth in the number of AI projects is far outpacing the increase in the number of deaf researchers in the same field. This situation differs from what sign linguistics one faced (and perhaps still does) regarding a similar lack of representation, as the development (and funding) in that field progressed at a slower pace. In contrast, AI is advancing at an unprecedented rate.

**Discussion and conclusion**

This paper examined the complex interplay of AI language technologies in the context of deaf people, sign language interpreting, and linguistic access. It highlights the interdependencies underpinning these access frameworks, the trade-offs inherent in these technologies, and the critical need to navigate these shifts carefully. The integration of AI language technologies into deaf accessibility practices brings opportunities and challenges, revealing the tension between innovation and the preservation of hard-won linguistic and accessibility rights. The tensions expressed by the European attendee in the introduction reflect one of the central challenges of this moment: navigating the promises of AI as a potential liberatory tool without undermining deaf communities' hard-won linguistic rights and access frameworks.

A key issue identified in the article is the problematic framing of AI sign language technologies as a substitute for human interpreters. This framing perpetuates a benchmark that undermines the very purpose of these technologies, which are not intended to compete with human interpreters but to support and reflect the diverse languaging practices of deaf communities. At its core, the development of these tools intersects with broader inclusion ideologies centred on interpreting, often reinforcing access hierarchies and introducing new forms of linguistic subordination to technology. These dynamic raises critical questions about not only how these tools are used but also how they redistribute access and reshape social hierarchies.

The concerns from deaf users, sign language interpreters and representative deaf organisations extend beyond who creates AI sign language technologies, to how these tools will be used once they are 'good enough' for real-world applications, and how this may lead to the potential erosion of hard-won linguistic and accessibility rights, particularly those rooted in human interpreting services. One major concern is that technological solutions risk being imposed on deaf people in situations where AI tools are not warranted, and the overestimation of AI capabilities might lead to compromised access. Although current AI sign language technologies are mainly targeted at so-called "low-stakes" contexts like public transportation (David & Bouillon, 2018; van Gemert et al., 2022) and hospitality (Leeson et al., 2024), developers are also pursuing "higher-stakes" areas like healthcare (Esselink et al., 2023) and emergency services (Guo et al., 2023; Martin et al., 2013). This shift, if not carefully managed, may exacerbate existing inequalities and undermine trust in these AI systems. These concerns do not come from thin air, considering the cost-effective approach often pursued by governments to address linguistic diversity and the “slippery slope” of access rights in deaf communities. The growing reliance on AI tools in low-stakes contexts, such as public transportation and customer services, might initially be seen as acceptable. However, these small adaptations could expand into higher-stakes contexts like healthcare or legal proceedings, where the stakes for deaf people are much higher. At the same time, policy measures can help but, in some ways, also hinder the implementation of a human-in-command approach. For instance, the EU Accessibility Act, set to take effect in June 2025, requires public and private entities, including broadcasters, to ensure their services and content are accessible to people with disabilities, covering measures like subtitles, audio descriptions, and sign language translation. The “shortage of human sign language interpreters”, perceived or real, and the legal obligation to meet accessibility requirements is frequently used here as a justification for developing sign language avatars for live broadcasting, despite deaf community preferences. This dynamic again highlights the interplay of technoableism and the interconnected discourses surrounding technology and sign language interpreting, where technological solutions are imposed as a quick fix for systemic challenges, without addressing the root causes of inequality. Deaf-led research in sign language AI is essential to challenging these power dynamics, ensuring ethical practices, and preventing harm caused by misaligned technological solutions. This research must emphasise the collective nature of access, emphasising choice, agency, and human oversight to ensure alignment with the diverse desires and preferences of deaf communities.

AI language technologies are not inherently disruptive but also generative, offering practical and theoretical opportunities. In a practical sense, they may alleviate the strain on interpreting services, expand access options, and empower deaf users by promoting AI literacy. Resilience lies not only in the capabilities of these tools but also in the capacity of deaf communities to adapt, innovate, and co-create alternative forms of access that reflect their values, needs, and aspirations. Theoretically, resilience lies in the development of new theoretical and methodological frameworks in Deaf Studies, Sign Language Interpreting Studies and sign language (socio)linguistics so they remain relevant and equitable in the AI era.

As highlighted in the introduction, technological shifts rarely occur overnight but instead embed themselves quietly into systems and practices. By the time their effects are fully visible, they may have already reshaped the accessibility landscape in profound and irreversible ways—both positively and negatively. Engaging critically with these shifts as they unfold is essential to fostering AI systems that are not only innovative and cutting-edge but also equitable, inclusive, and trustworthy, and serve rather than subordinate linguistic practices to technological imperatives.